\documentclass{aa}
\usepackage{graphicx}

\normalfont

\def\etal{{\it et~al. }}
\def\J2{$J_{2\odot}$}
\def\cmr2{C/MR$^{2}$}
\def\c22{$C_{22}$}
\def\c20{$C_{20}$}
\def\32{$3$:$2$}

\begin{document}

     \title{Mercury's spin-orbit model and signature of its dynamical
     parameters}
     \author{N. Rambaux and E. Bois }
     \offprints{"rambaux@obs.u-bordeaux1.fr"}
     \institute{Observatoire Aquitain des Sciences de l'Univers,
                     UMR CNRS/INSU 5804 (L3AB), B.P. 89, F-33270,
                    Floirac, France}
     \date{Received ; accepted }

\abstract{
The 3:2 spin-orbit resonance between the rotational and orbital motions of Mercury results from 
a functional dependance on a tidal friction adding to a non-zero eccentricity with a permanent
asymmetry in the equatorial plane of the planet. The upcoming space missions, MESSENGER and 
BepiColombo with onboard instrumentation capable of measuring the Mercury's rotational parameters,
stimulate the objective to attempt to an accurate theory of the planet's rotation. We have used our BJV relativistic 
model of solar system integration including the spin-orbit motion of the Moon. This model had been previously 
built in accordance with the requirements of the Lunar Laser Ranging observational accuracy. We extended
this model to the spin-orbit couplings of the terrestrial planets including Mercury; the updated model is called 
SONYR (acronym of Spin-Orbit N-BodY Relativistic model). An accurate rotation of Mercury has been then
obtained. Moreover, the conception of the SONYR model is suitable for analyzing the different families of 
hermean rotational librations. We accurately identify the non-linear relations between the rotation of Mercury and 
its dynamical figure (\cmr2, $C_{20}$, and $C_{22}$). Notably, for a variation of 1\% on the \cmr2 value,
signatures in the $\varphi$ hermean libration in longitude as well as in the $\eta$ obliquity of the planet are
respectively 0.45 arcseconds (as) and 2.4 milliarcseconds (mas). These determinations provide new 
constraints on the internal structure of Mercury to be discussed with the expected accuracy forecasted
in the BepiColombo mission (respectively 3.2 and 3.7 as according to Milani \etal 2001). 
}

     \maketitle

\section{Introduction} 

Pettengill and Dyce discovered in 1965 the 3:2 spin-orbit resonance state of Mercury 
(the rotational and orbital periods are 56.646 and 87.969 days respectively). This particular resonance results from 
a functional dependance on a tidal friction adding to a non-zero eccentricity with a permanent asymmetry in the 
equatorial plane of the planet. Besides, in a tidally evolved system, the spin pole is expected to be trapped in a 
Cassini state: the orbital and rotational parameters are indeed matched in such a way that the spin pole, the 
orbit pole, and the solar system invariable pole remain coplanar while the spin and orbital poles on average 
precess at the same rate (Colombo 1965; Peale 1969, 1973). 

The upcoming missions, MESSENGER (Solomon \etal 2001) and BepiColombo (Milani \etal 2001) with onboard 
instrumentation capable of measuring the rotational parameters of Mercury stimulate the objective to 
attempt to an accurate theory of the Mercury's rotation. The current method for obtaining constraints on
the state and structure of the hermean nucleus is based on the assertions introduced by Peale in 1976. 
According to this author, the determination of the four parameters, namely $C_{20}$, $C_{22}$, $\eta$ and 
$\varphi$, should be sufficient to constrain size and state of the Mercury's core ($C_{20}$ and $C_{22}$ are
spherical harmonics of the second degree, $\eta$ is the obliquity of Mercury, and $\varphi$ is the libration in 
longitude of 88 days).

In the first section of this paper, we present our approach which takes into account the coupled spin-orbit
motion such as it derives from the integration of the entire solar system. In the second section, we present the
dynamical behavior of the Mercury's spin-orbit motion. In the last section, following our previous analysis of the
different families of hermean rotational librations (see our previous paper Rambaux and Bois 2003), we make
in evidence some accurate signatures due to parameters related to the dynamical figure of Mercury 
(\cmr2, $C_{20}$, and $C_{22}$) in the librations as well as in the obliquity of the planet. We find 
0.45 arcseconds (as) and 2.4 milliarcseconds (mas) respectively on $\varphi$ and $\eta$ for a variation of 1\%
on the \cmr2 value. These determinations provide new constraints on the internal structure of Mercury to be
discussed with the expected accuracy forecasted in the BepiColombo mission (respectively 3.2 and 3.7 as 
according to Milani \etal 2001).  

\section{The Spin-Orbit \textit{N}-Body Relativistic model} 

For obtaining the real motion of Mercury, we have used our BJV model of solar system integration including the 
coupled spin-orbit motion of the Moon. This model, expanded in a relativistic framework, had been previously 
built in accordance with the requirements of the Lunar Laser Ranging observational accuracy (Bois 2000; Bois
\& Vokrouhlick\'y 1995). We extended the BJV model by generalizing the spin-orbit couplings to the terrestrial
planets (Mercury, Venus, Earth, and Mars).
The model is at present called SONYR (acronym of Spin-Orbit N-BodY Relativistic model). As a consequence, the 
SONYR model gives an accurate simultaneous integration of the spin-orbit motion of Mercury. The integration of 
the solar system, including the Mercury's spin-orbit motion, uses a global reference system given by the solar 
system barycenter. The model is solved by modular numerical integration and controlled in function of the different 
physical contributions and parameters taken into account. It permits the analysis of the different families of librations 
and the identification of their causes (such as the planetary interactions) or their interdependences with the parameters
involved in the dynamical figure of the planet. A complete description of the SONYR model is given in our previous 
paper (Rambaux and Bois 2003). The spin-orbit motion of Mercury is characterized by two proper frequencies 
(namely $\Phi$ = 15.847 years and $\Psi$ = 1066 years) and its 3:2 resonance presents a second synchronism, 
which can be understood as a \textit{spin-orbit secular resonance} ($\Pi$ = 278\,898 years).

Using the SONYR model and its method of analysis (modular and controlled numerical integration,
differentiation method), the librations are accurately surrounded and identified. 
Our previous paper Rambaux and Bois (2003) presents a detailed analysis of the whole
spin-orbit mechanism of Mercury as well as its main librations.

\section{The Mercury's spin-orbit motion}

Using our SONYR model, the present work is devoted for studiing the impact of the dynamical figure
of Mercury on its spin-orbit motion. We have then analyzed the impact of the three dynamical parameters 
\cmr2, $C_{20}$, and $C_{22}$ on the rotational motion of Mercury.

\begin{figure}[!ht]
      \begin{center}
        \hspace{-0.55cm}
       \includegraphics[width=6cm]{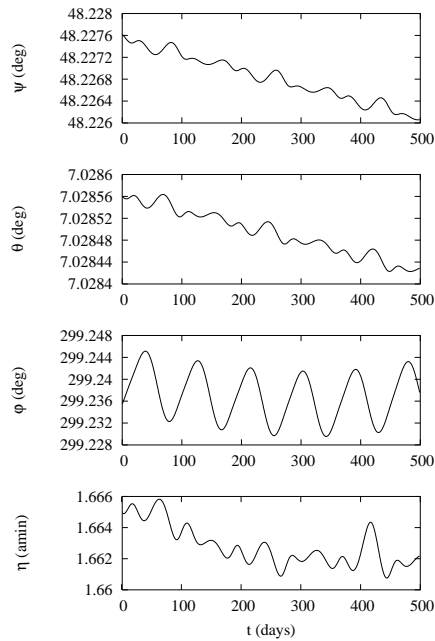}
        \caption{The rotational motion of Mercury expressed in the 
        classical 3-1-3 Eulerian sequence ($\psi$, $\theta$, 
        $\varphi$) with respect to the ecliptic reference frame.
        In the bottom panel is plotted the obliquity $\eta$. The 
        four curves are plotted over 500 days (horizontal axes). 
        Degrees are on the vertical axes for $\psi$, $\theta$, and
        $\varphi$, while the vertical axis for $\eta$ is plotted in arcminutes (amin).}
        \label{fig1}
      \end{center}
\end{figure}
        
Figure~\ref{fig1} presents the dynamical behavior of the Mercury's rotation expressed
by the Euler angles $\psi,\theta$, and $\varphi$ (three first panels) as well as its obliquity $\eta$ 
plotted in the fourth panel over 500 days. The Euler angles ($\psi,\theta,\varphi$) related 
to the 3-1-3 angular sequence describe the evolution of the body-fixed axes $Oxyz$ with
respect to the axes of the local reference frame $OXYZ$. 
Let us recall the definition used for these angles~: $\psi$ is the precession
angle of the polar axis $Oz$ around the reference axis $OZ$, $\theta$ is the
nutation angle representing the inclination of $Oz$ with respect to $OZ$, and 
$\varphi$ is the rotation around $Oz$ and conventionally understood as the rotation
of the greatest energy (it is generally called the proper rotation). The axis of
inertia around which is applied the proper rotation is called the axis of figure
and defines the North pole of the rotation (Bois 1992). Let us remark that in
this Figure, we have removed the mean rotation of $58.646$ days in the
$\varphi$ angle in order to better distinguish the librations. The resulting amplitude
of libration appearing in $\varphi$ is 20 as, which is in good agreement with
Balogh and Giamperi (2002).

The integration of the SONYR model including the simultaneous spin-orbit 
motion of Mercury gives also the dynamical behavior of the hermean obliquity
by the way of the following relation~:
\begin{equation}
    \cos \eta = \cos i \cos\theta  + \sin i \sin\theta \cos(\Omega -\psi)
    \label{eq:obl}
\end{equation}
where $i$ and $\Omega$ are respectively the inclination and ascending node 
of the Mercury's orbital plane relative to the ecliptic plane (the obliquity is the
angle between the polar axis and the orbital plane).
In our computations presented in this paper, the used initial conditions 
are listed in Table~\ref{paraci}. For $\eta$, $\dot{\psi}$, and $\dot{\theta}$, we
use the values determined in our previous work (Rambaux and Bois 2003).  

\begin{figure*}[!ht]
      \begin{center}
        \includegraphics[width=10cm,angle=-90]{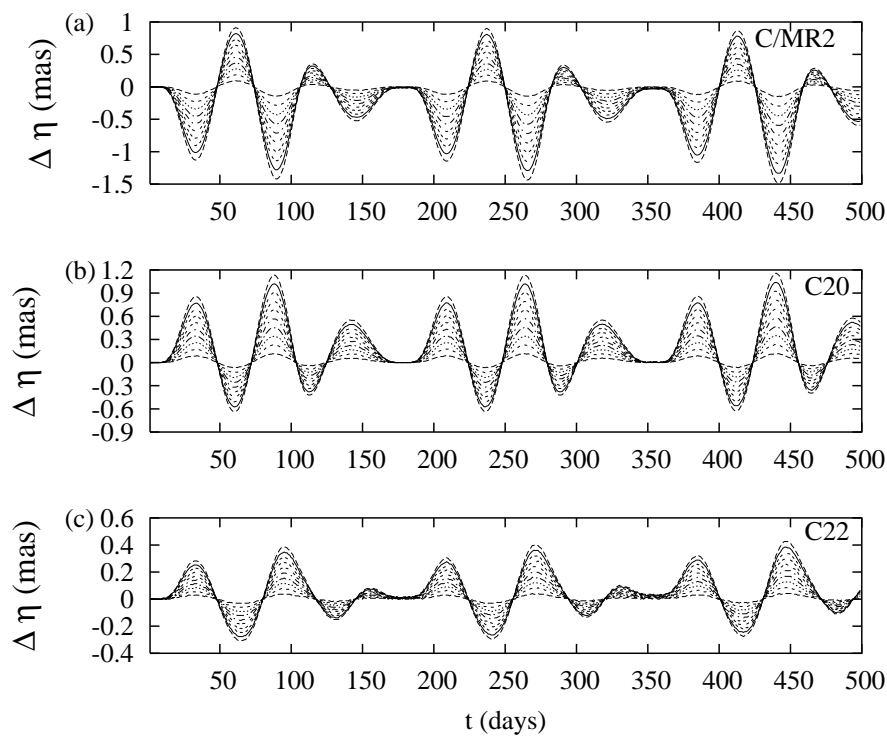}
        \caption{Signatures of the variations (with a step equal to 0.1\%) of the dynamical 
        parameter values on the Mercury's obliquity: (a) \cmr2, (b) $C_{20}$, and (c) $C_{22}$. 
        On each panel, 10 curves are plotted over 500 days; the maximum of amplitude is obtained for 
        1\%. Milliarcseconds (mas) are on the vertical axes.}
        \label{figobl}
      \end{center}
\end{figure*}

The dynamical structure of the Mercury's rotation given in Figure~\ref{fig1} 
results from the direct solar torque whose coordinates include the planetary 
interactions. The effect of the gravitational torque due to Venus is 
five orders of magnitude smaller than the one due to the Sun. The $P_{\varphi}$ rotation 
period of 58.646 days appears in the $\psi$ and $\theta$ angles, as well as in $\eta$. 
Signature of the $P_{\lambda}$ orbital period of 87.969 days is clearly visible 
in the $\varphi$ angle (this angle is called in literature the libration in longitude of 
88 days). A third period appears in the $\psi$, $\theta$, and $\eta$ angles, namely 
176.1 days. This period $\widetilde{P}$ results from the \32  spin-orbit resonance 
($\widetilde{P} = 2 P_{\lambda}=3 P_{\varphi}$).

\begin{table}[!htb]
      \centering
       \caption{Our initial conditions at 07.01.1969 (equinox 
       J2000): (a) Mean values derived from the SONYR model;
       (b) Seidelmann \etal (2002). In addition, our SONYR value 
       of the Mercury's obliquity is 1.6 amin.}
      \begin{tabular}{rclrcl}
         \hline    
           \hline      
           \multicolumn{6}{c}{Mercury } \\
       \multicolumn{6}{c}{ Rotation angles}\\
           \hline
          & & & &  \\	   
       $ \psi     $ &  = &   $48.386$~deg    & 
       $\dot \psi $ & = & -0.616 10$^{-7}$~deg/day (a) \\
       $ \theta  $& =   & $ 7.031$~deg    &
       $\dot \theta$ &= &   -0.267 10$^{-8}$~deg/day (a)  \\
       $\varphi $& = &$299.070$~deg    & 
       $\dot \varphi$& =& 6.138505~deg/day (b) \\
              \hline
       \end{tabular}
    \label{paraci}
\end{table}

\section{The impact of the dynamical parameters}

\begin{figure*}[!ht]
      \begin{center}
        \includegraphics[width=10cm,angle=-90]{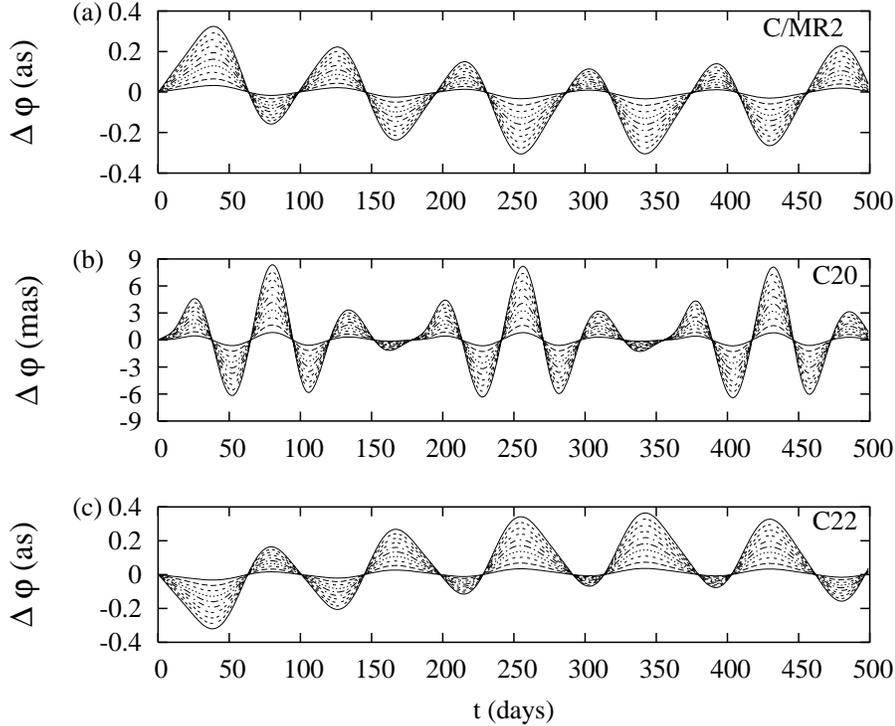}
        \caption{Signatures of the variations (with a step equal to 0.1\%) of the dynamical 
        parameter values on the Mercury's libration in longitude: (a) \cmr2, (b) $C_{20}$, 
        and (c) $C_{22}$. 
        On each panel, 10 curves are plotted over 500 days; the maximum of amplitude is
        obtained for 1\%. Arcseconds (as) are on the vertical axes 
        of the top and bottom panels while it is about milliarcseconds 
        (mas) in the middle panel.}
        \label{figphi}
      \end{center}
\end{figure*}

The gravity field of Mercury is globally unknown. The tracking 
data from the three fly-by of Mariner 10 in 1974-75 have been 
re-analyzed by Anderson \etal (1987) in order to give an accurate
estimation of the normalized coefficients $C_{20}$ and $C_{22}$. 
The resulting nominal values are  \cmr2=0.34, $C_{20}=(6.0 \pm 2.0)\; 10^{-5}$, 
and $C_{22}=(1.0 \pm 0.5)\; 10^{-5}$. As a consequence, the dynamical coefficients 
$\alpha=(C-B)/A$, $\beta=(C-A)/B$, and $\gamma=(B-A)/C$ (where $A,B$, 
and $C$ are the three principal moments of inertia of Mercury) are infered from $C_{20}$
and $C_{22}$ by the following formulae~:
\begin{eqnarray}
    \alpha & = & \frac{-C_{20}-2C_{22}}{C_{20}-2C_{22}+C/MR^{2}} \nonumber  \\ 
    \beta & = & \frac{-C_{20}+2C_{22}}{C_{20}+2C_{22}+C/MR^{2}}      \\
    \gamma & = & \frac{4C_{22}}{C/MR^{2}}    \nonumber \\  \nonumber 
\end{eqnarray}
In these conditions, we may assume the uncertainties on the Mercury's dynamical 
parameter values as dynamical variations on the Mercury's rotation. We have
computed the impact of the variations of the greatest principal moment of inertia,
\cmr2, as well as $C_{20}$ and $C_{22}$, both on the obliquity and the libration in 
longitude. Figures~\ref{figobl} and \ref{figphi} present the results plotted over 500 days. 

These Figures show three panels corresponding to signatures of the 
variations of \cmr2 (a), $C_{20}$ (b), and $C_{22}$ (c) on $\eta$ and $\varphi$.  
In each panel, 10 integrations with a step of 0.1\% over the given 
parameter are plotted over 500 days; the maximum of amplitude is obtained for 1\%.
In each plot, the maximum amplitude is reached with the characteristic 
period of $176.1$ days for the obliquity $\eta$, and 87.969 days for the 
libration in longitude $\varphi$. The results related to variations of 
1\% are listed in Table~\ref{tbl:res}.

\begin{table}[!htb]
      \centering
       \caption{Results of the 1\% variations of the dynamical 
        parameters on the Mercury's obliquity and libration in 
        longitude.}
      \begin{tabular}{cccc}
         \hline    
           \hline     
              & $C/MR^{2}$ & $C_{20}$ & $C_{22}$   \\
	\hline
          & & &  \\	  
          $\eta$ & 2.4 mas & 1.8 mas & 0.7 mas  \\ 
          $\varphi$ & 0.45 as & 14 mas &  0.45 as  \\   
              \hline
       \end{tabular}
    \label{tbl:res}
\end{table}

In the Figure~\ref{figobl}, a variation of 1 \% on the \cmr2 parameter induces
a peak to peak amplitude of 2.4 mas in the obliquity, whereas with the $C_{20}$ 
and $C_{22}$ coefficients, the amplitudes are 1.8 and 0.7 mas respectively. 
The characteristic period of these signatures is $\widetilde{P}=176.1$ 
days, which expresses the spin-orbit coupling of Mercury. The 
dynamical behavior of $\eta$ proves to be the right tracer for 
analyzing the relationship between the rotation of Mercury and its 
dynamical figure.

In the Figure~\ref{figphi}, the impact of variations of the $C_{20}$ 
value on the libration in longitude is 30 times smaller than the one 
due to variations of \cmr2 and $C_{22}$. This is make clear by the 
fact that the dynamical parameter $\gamma$ depends only on $C_{22}$ and 
\cmr2. As a consequence, $C_{20}$ has only an indirect action on the $\varphi$
angle. In addition, the dynamical behaviors in Figure~\ref{figphi}, panels (a) and
(c), are symmetrical; this fact reflects structure of the third equation in the formulae (2). 
The relation between $C_{20}$ and $\varphi$ proves to be a good tracer of the $\widetilde{P}$
spin-orbit period while the relations between \cmr2 and $\varphi$, or $C_{22}$ and $\varphi$, 
express the $P_{\lambda}$ period.

\section{Conclusion}

One important objective of the upcoming space missions, MESSENGER and BepiColombo, 
is to constrain the nature of the Mercury's core. The theoretical framework follows the procedure
proposed by Peale (1976) and Peale \etal (2002). The onboard instrumentations are planed for
measuring the Mercury's rotational parameters with a great accuracy.
However, our work shows that for obtaining an excepted accuracy up to 1\% on the \cmr2 coefficient,
the measurements have to reach at least 0.45 as on the Mercury's rotation and 2.4 mas on the obliquity. 
To our knowledge, these high accuracies are not yet achieved within the simulations.

The Mercury's rotation strongly depends on the \cmr2 (0.45 as) and $C_{22}$ (0.45 as) coefficients.
Besides, the obliquity of Mercury hierarchically depends on  \cmr2 (2.4 mas), $C_{20}$ (1.8 mas), and 
$C_{22}$ (0.7 mas). The dynamical behavior of Mercury's obliquity is finally the tracer for analyzing 
the non-linear relations between the rotation of Mercury and parameters of its dynamical figure.

 \begin{acknowledgements}
The authors thank Anne Lema\^{\i}tre and Sandrine D'Hoedt for useful discussions.
 \end{acknowledgements}
   

\end{document}